\documentclass[rnaas]{aastex63}
\pdfoutput=1 
\usepackage{amsmath,amstext}
\usepackage[T1]{fontenc}
\usepackage[figure,figure*]{hypcap}
\usepackage[version=4]{mhchem} 

\shorttitle{The Search for MeV Electrons with {\sl New Horizons}}
\shortauthors{Keeney et~al.}

\begin{document}

\title{The Search for MeV Electrons 2-45~AU from the Sun with the Alice Instrument Microchannel Plate Detector Aboard {\sl New Horizons}}

\author[0000-0003-0797-5313]{B. A. Keeney}
\affiliation{Southwest Research Institute, Boulder, CO 80302, USA}
\email{keeney@boulder.swri.edu}
\author{M. Versteeg}
\affiliation{Southwest Research Institute, San Antonio, TX 78238, USA}
\author{J. Wm. Parker}
\affiliation{Southwest Research Institute, Boulder, CO 80302, USA}
\author{S. A. Stern}
\affiliation{Southwest Research Institute, Boulder, CO 80302, USA}
\author{P. Brunts}
\affiliation{Stellar Solutions, Inc., Palo Alto, CA 94306, USA}
\author{M. W. Davis}
\affiliation{Southwest Research Institute, San Antonio, TX 78238, USA}
\author{H. A. Elliott}
\affiliation{Southwest Research Institute, San Antonio, TX 78238, USA}
\author{K. Ennico}
\affiliation{NASA Ames, Moffett Field, CA 94035, USA}
\author{G. R. Gladstone}
\affiliation{Southwest Research Institute, San Antonio, TX 78238, USA}
\affiliation{University of Texas at San Antonio, San Antonio, TX 78249, USA}
\author{R. L. McNutt, Jr.}
\affiliation{JHU APL, Laurel, MD 20723, USA}
\author{C. B. Olkin}
\affiliation{Southwest Research Institute, Boulder, CO 80302, USA}
\author{K. D. Retherford}
\affiliation{Southwest Research Institute, San Antonio, TX 78238, USA}
\affiliation{University of Texas at San Antonio, San Antonio, TX 78249, USA}
\author{K. N. Singer}
\affiliation{Southwest Research Institute, Boulder, CO 80302, USA}
\author{J. R. Spencer}
\affiliation{Southwest Research Institute, Boulder, CO 80302, USA}
\author{A. J. Steffl}
\affiliation{Southwest Research Institute, Boulder, CO 80302, USA}
\author{H. A. Weaver}
\affiliation{JHU APL, Laurel, MD 20723, USA}
\author{L. A. Young}
\affiliation{Southwest Research Institute, Boulder, CO 80302, USA}

\begin{abstract}
    The Alice UV spectrograph aboard NASA's {\sl New Horizons} mission is sensitive to MeV electrons that penetrate the instrument's thin aluminum housing and interact with its microchannel plate detector. We have searched for penetrating electrons at heliocentric distance of 2-45 AU, finding no evidence of discrete events outside of the Jovian magnetosphere. However, we do find a gradual long-term increase in the Alice instrument's global dark count rate at a rate of $\sim1.5$\% per year, which may be caused by a heightened gamma-ray background from aging of the spacecraft's radioisotope thermoelectric generator fuel. If this hypothesis is correct, then the Alice instrument's global dark count rate should flatten and then decrease over the next 5-10 years.
\end{abstract}

\section{}

When NASA's {\sl New Horizons} spacecraft flew past Jupiter for a gravity assist in 2007, we  discovered that the UV spectrograph, Alice \citep{stern08}, is sensitive to electrons with energies $>1$~MeV, which penetrate the instrument's thin aluminum housing and interact with its microchannel plate detector \citep{steffl12}. Subsequently, the Alice instrument team regularly scheduled observations in High Energy Electron Test (HEET) mode to search for MeV electrons elsewhere in the solar system. HEET measurements were obtained when the Alice aperture door was closed and the high-voltage power was on, excluding times when the spacecraft was near Jupiter or when the Sun was in the solar occultation aperture. This same filtering was retroactively applied to identify suitable measurements obtained throughout the mission flight. 

Alice acquired 352 sequences containing 181,567 HEET measurements, where a sequence is a continuous period of HEET measurements during which the underlying instrument properties are constant, and a HEET measurement is a ``dark'' global count rate taken at either 1- or 30-second cadence. HEET sequences were collected between May 2006 and September 2019 at heliocentric distances of 2-45~AU. Alice obtained HEET data during 66\% of the time in which its high-voltage supply was powered on; however, the Alice high-voltage supply has only been powered on for $\sim1$\% of the mission to date. This creates a challenge when comparing to other {\sl New Horizons} particle detection instruments; e.g., SWAP \citep[$20~\mathrm{eV}<E<7.5~\mathrm{keV}$;][]{mccomas08} and PEPSSI \citep[$E<1$~MeV;][]{mcnutt08}, collect data nearly continuously, so they have a higher likelihood of detecting discrete changes in the spacecraft's particle environment than Alice, albeit at lower energies.

\begin{figure}
  \fig{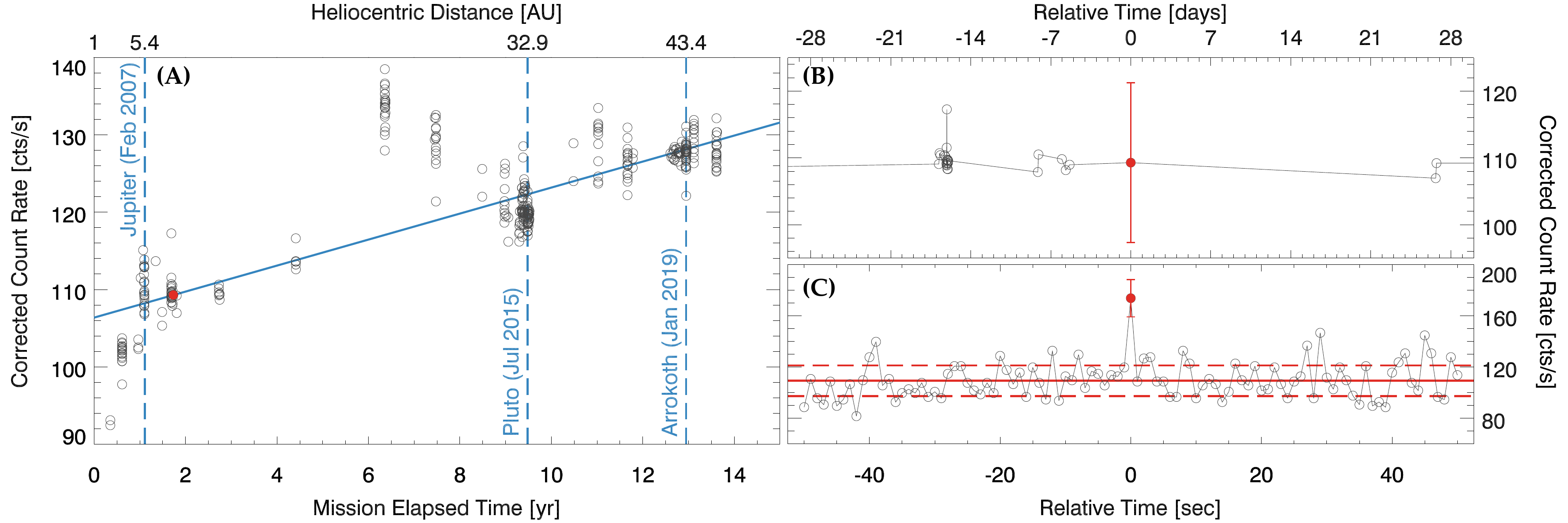}{\columnwidth}{}
  \vspace{-3em}
  \caption{Alice count rate variations at different time scales. Panel~A shows the gradual long-term increase at a rate of $\sim1.5$\% per year. Panel~B shows all sequences taken within 1~month of the sequence highlighted in Panel~A; the variations on this time scale are random and within the statistical noise. Panel~C shows the most statistically significant outlier in the dataset compared to a subset of measurements in the same sequence. The outlier represents a $5.4\sigma$ deviation above the mean and belongs to the sequence highlighted in Panels~A and B; the horizontal lines show the sequence mean and standard deviation from Panel~B.
    \label{fig:met}}
\end{figure}

HEET measurements were corrected to remove instrumental ``stim pulse'' events and correct for high-voltage changes and detector temperature variations. \autoref{fig:met} (Panel~A) shows the mean count rate of each sequence, which increases gradually over time. The cause of this increase is unclear. SWAP and PEPSSI do not observe similar trends in the spacecraft's particle environment, nor can this trend be explained by aging of the instrument. 

It is conceivable that this long-term increase is caused by changes in the spacecraft's relativistic particle environment that are undetectable by other instruments. However, the flux of relativistic electrons from Jupiter \textit{decreases} at $\sim2$\% per AU at $R>11$~AU \citep{eraker82}, and galactic cosmic rays, mostly protons, do not dominate the relativistic particle environment until larger heliocentric radii \citep{stone13,webber18}. Another hypothesis is that the gradual increase in Alice count rate reflects a heightened gamma-ray background from aging of the spacecraft's radioisotope thermoelectric generator fuel \citep{cockfield06}. If this hypothesis is correct, then the Alice HEET count rate should flatten and then decrease over the next 5-10 years \citep{rinehart01}.

Examining each HEET sequence individually, we find occasional high-$\sigma$ outliers, but none are associated with a coherent increase in count rate lasting more than 2-3~measurements. The most statistically significant outlier ($5.4\sigma$ above the sequence mean)\footnote{Assuming Gaussian statistics, a dataset with 181,567 measurements has a 3\% chance of one outlier at the $5.4\sigma$ level.} is shown in \autoref{fig:met} (Panel~C). Although the count rate is high there is no correlation with adjacent HEET measurements, or with contemporaneous SWAP or PEPSSI measurements, suggesting that there is no physical cause for this anomalous value. Examining all high-$\sigma$ outliers individually, there is no correlation between their occurrence in the Alice HEET sequences and increases in concurrent SWAP and PEPSSI particle detection rates or penetrating radiation events, unlike what was seen in the the high-energy-electron rich environment of the Jovian magnetosphere \citep{steffl12}.

\acknowledgments
We thank the {\sl New Horizons} operations team for finding the interstitial moments to schedule HEET mode, and the Alice team, including our late team member David C. Slater, for an excellent instrument. We also acknowledge Sensor Sciences, LLC for assistance understanding the long-term behavior of the Alice detector. This work was supported by NASA's {\sl New Horizons} project.

\newpage
\bibliographystyle{aasjournal}
\bibliography{references}

\begin{thebibliography}{}
\expandafter\ifx\csname natexlab\endcsname\relax\def\natexlab#1{#1}\fi
\providecommand{\url}[1]{\href{#1}{#1}}

\bibitem[{Cockfield(2006)}]{cockfield06}
Cockfield, R. 2006, Preparation of RTG F8 for the Pluto New Horizons Mission.
\newblock \url{https://arc.aiaa.org/doi/abs/10.2514/6.2006-4031}

\bibitem[{{Eraker}(1982)}]{eraker82}
{Eraker}, J.~H. 1982, \apj, 257, 862

\bibitem[{{McComas} {et~al.}(2008){McComas}, {Allegrini}, {Bagenal}, {Casey},
  {Delamere}, {Demkee}, {Dunn}, {Elliott}, {Hanley}, {Johnson}, {Langle},
  {Miller}, {Pope}, {Reno}, {Rodriguez}, {Schwadron}, {Valek}, \&
  {Weidner}}]{mccomas08}
{McComas}, D., {Allegrini}, F., {Bagenal}, F., {et~al.} 2008, \ssr, 140, 261

\bibitem[{{McNutt} {et~al.}(2008){McNutt}, {Livi}, {Gurnee}, {Hill}, {Cooper},
  {Andrews}, {Keath}, {Krimigis}, {Mitchell}, {Tossman}, {Bagenal}, {Boldt},
  {Bradley}, {Devereux}, {Ho}, {Jaskulek}, {Lefevere}, {Malcom}, {Marcus},
  {Hayes}, {Moore}, {Perry}, {Williams}, {Wilson}, {Brown}, {Kusterer}, \&
  {Vandegriff}}]{mcnutt08}
{McNutt}, R.~L., {Livi}, S.~A., {Gurnee}, R.~S., {et~al.} 2008, \ssr, 140, 315

\bibitem[{Rinehart(2001)}]{rinehart01}
Rinehart, G.~H. 2001, Progress in Nuclear Energy, 39, 305 .
\newblock
  \url{http://www.sciencedirect.com/science/article/pii/S0149197001000051}

\bibitem[{{Steffl} {et~al.}(2012){Steffl}, {Shinn}, {Gladstone}, {Parker},
  {Retherford}, {Slater}, {Versteeg}, \& {Stern}}]{steffl12}
{Steffl}, A.~J., {Shinn}, A.~B., {Gladstone}, G.~R., {et~al.} 2012, Journal of
  Geophysical Research (Space Physics), 117, A10222

\bibitem[{{Stern} {et~al.}(2008){Stern}, {Slater}, {Scherrer}, {Stone},
  {Dirks}, {Versteeg}, {Davis}, {Gladstone}, {Parker}, {Young}, \&
  {Siegmund}}]{stern08}
{Stern}, S.~A., {Slater}, D.~C., {Scherrer}, J., {et~al.} 2008, \ssr, 140, 155

\bibitem[{{Stone} {et~al.}(2013){Stone}, {Cummings}, {McDonald}, {Heikkila},
  {Lal}, \& {Webber}}]{stone13}
{Stone}, E.~C., {Cummings}, A.~C., {McDonald}, F.~B., {et~al.} 2013, Science,
  341, 150

\bibitem[{Webber(2018)}]{webber18}
Webber, W.~R. 2018, in Cosmic Rays, ed. Z.~Szadkowski (Rijeka: IntechOpen).
\newblock \url{https://doi.org/10.5772/intechopen.75877}

\end{thebibliography}

\end{document}